\documentclass[%
 reprint,
 amsmath,amssymb,
 aps,
prb,
]{revtex4-1}

\usepackage{graphicx}% Include figure files
\usepackage{dcolumn}% Align table columns on decimal point
\usepackage{bm}% bold math
\usepackage{color}

\begin{document}

\preprint{APS/123-QED}

\title{Photon assisted tunneling through three quantum dots with spin-orbit-coupling}% Force line breaks with \\

\author{Han-Zhao Tang}
\affiliation{College of Physical Science and Information Engineering and Hebei Advanced Thin Film Laboratory, Hebei Normal University, Shijiazhuang, Hebei 050024, People's Republic of China}

\author{Xing-Tao An}
\email{anxt@hku.hk}
\affiliation{School of Sciences, Hebei University of Science and Technology, Shijiazhuang, Hebei 050018, People's Republic of China}
\affiliation{Department of Physics and Center of Theoretical and Computational Physics, University of Hong Kong, Hong Kong, People's Republic of China}
\author{Ai-Kun Wang}
\affiliation{School of Sciences, Hebei University of Science and Technology, Shijiazhuang, Hebei 050018, People's Republic of China}

\author{Jian-Jun Liu}
\email{liujj@mail.hebtu.edu.cn}
\affiliation{Physics Department, Shijiazhuang University, Shijiazhuang, Hebei 050035, People's Republic of China}

\date{\today}% It is always \today, today,

\begin{abstract}
The effect of an ac electric field on quantum transport properties in a system of three quantum dots, two of which are connected in parallel while the third is coupled to one of the other two, is investigated theoretically. Based on the Keldysh nonequilibrium Green's function method, the spin-dependent current, occupation number and spin accumulation can be obtained in our model. An external magnetic flux, Rashba spin orbit coupling (SOC) and intradot Coulomb interactions are considered. The magnitude of the spin-dependent average current and the positions of the photon assisted tunneling (PAT) peaks can be accurately controlled and manipulated by simply varying the strength of the coupling and the frequency of the ac field. A particularly interesting result is the observation of a new kind of PAT peak and a multiple electron-photon pump effect that can generated and controlled by the coupling between the quantum dots. In addition, the spin occupation number and spin accumulation can be well controlled by the Rashba SOC and the magnetic flux.
\begin{description}
\item[PACS numbers]
85.35.-p, 32.80.-t, 71.70.Ej, 85.75.-d
\end{description}
\end{abstract}

\pacs{Valid PACS appear here}% PACS, the Physics and Astronomy
\maketitle

%\tableofcontents

\section{\label{sec:level1}Introduction}

Electron transport through low dimensional nanostructures to which a microwave (MW) field is applied has received increased attention in recent years. An important characteristic of these systems is that the electron in the system can exchange an energy $n\hbar\omega$ with the external fields, where $n=\pm1,\pm2,\ldots$, and $\omega$ is the frequency of the external field, thus leading to several new inelastic tunneling channels. This phenomenon has been called the photon assisted tunneling (PAT) effect.

The effects of a MW field on superconductivity were investigated by Tien $et~al$.\cite{1} in the 1960s. Later, different theoretical methods were proposed, such as the time-dependent Schr\"{o}dinger equation,\cite{2,3,4} the transfer Hamiltonian method,\cite{5,6} the Master equation \cite{7,8} and the Keldysh nonequilibrium Green's function method.\cite{9,10,11,12,13,14} Experimentally, the PAT effect has been observed in quantum dot (QD) systems with a single QD,\cite{15} and in a system with double QDs.\cite{16,17,18} The observation of the photon-electron pump phenomenon in a QD system which is controlled by an ac field has been reported by Kouwenhoven $et~al$.\cite{19,20} Sun $et~al$. have investigated electron tunneling through a QD\cite{21}and in a quantum-dot-molecule\cite{22} irradiated by a MW field. Besides the single QD system, time-dependent tunneling through double\cite{23,24,25} and triple\cite{26,27} coupled QDs have also received great attention both experimentally and theoretically, in many cases because of the potential applications in quantum computing devices.
\

When a device is prepared in a semiconductor with a perpendicular electric field, Rashba spin-orbit coupling (SOC) will appear in the system, which leads to a nonzero spin-dependent phase $\sigma_{R}$.\cite{28} In addition, the time reversal symmetry can be broken by a magnetic flux $\varphi$. If both these effects are present, the average current is expected to become spin polarized. L\"{u} $et~al$.\cite{26} have proposed a spin filter using a triple QD system with dc bias. However, to the best of our knowledge, little attention has been paid to spin transport and the occupation number in such device in a MW field, especially a system with Rashba SOC. In order to study the impact of coupling between QDs in a device with three QDs, we have constructed a theoretical model to investigate the PAT effect and photon-electron pump phenomenon when the electron-electron ($e-e$) interaction, Rashba SOC and an external magnteic field are all considered.

In this paper, using the Keldysh nonequilibrium Green's function method, we analytically solve for the time-dependent current through two QDs connected in parallel with a side-coupled QD, the whole system being irradiated by a MW field. This paper is organized as follows. The model and analytic method are introduced in Sec. \uppercase \expandafter {\romannumeral 2}. In Sec. \uppercase \expandafter {\romannumeral 3}, we discuss our results including the spin-dependent average current, the occupation numbers and the spin accumulation for various cases. Finally, a symmary is given in Sec. \uppercase \expandafter {\romannumeral 4}.

\section{\label{sec:level2}Model and formulation}
\begin{figure}
\centering
\includegraphics[scale=0.5]{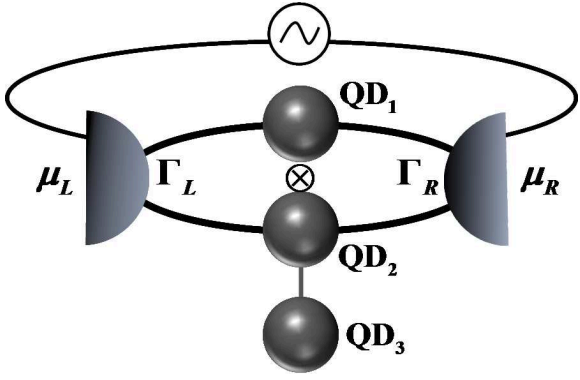}
\caption{\label{fig_1}Schematic diagram of a device consisting of two QDs connected in parallel with a third side-coupled QD connected to one of the other QDs. An ac bias is applied across the leads.}
\end{figure}
As shown in Fig. 1, the system we propose is composed of three QDs, which can also be seen as two QDs connected in parallel with a third side-coupled QD. The third QD is not directly coupled to the leads under ac bias. The Hamiltonian of the system can be described as:
\begin{equation}
  H=\sum_{\beta=L,R}H_{\beta}+H_{D}+H_{T}.\label{Eq1}
\end{equation}

The first term of the Hamiltonian in Eq. (1) describes the lead system:
\begin{equation}
  H_{\beta}=\sum_{k,s}\varepsilon_{\beta k}(t)\alpha^{\dag}_{\beta ks}\alpha_{\beta ks},\label{Eq2}
\end{equation}
where $\alpha^{\dag}_{\beta ks}$($\alpha_{\beta ks}$) is the creation (annihilation) operator of an electron with spin $s$ ($s=\uparrow,\downarrow$)and Bloch wave vector $k$ in the $\beta$ lead. The $\beta$ leads are the left lead and the right lead in the system. The electron energy $\varepsilon_{\beta k}(t)=\varepsilon_{\beta k}^{0}+eV+W_{\beta}(t)=\varepsilon_{\beta k}^{0}+eV_{\beta}-eW_{\beta}\cos(\omega t)$. Here, $\varepsilon_{\beta k}^{0}$ is a single particle energy, $ V_{\beta}$ is a dc bias (electron charge $-e$) and the ac bias of frequency $\omega$ is given by $W_{\beta}\cos(\omega t)$. The effect of ac fields influencing the energy levels of the source and the drain was studied by Jauho $et~al$..\cite{11}

The second term in Eq. (1) describes the QD system:
\begin{equation}
  H_{D}=\sum_{s,i=1,2,3}\varepsilon_{i}(t)d^{\dag}_{is}d_{is}-(td^{\dag}_{2s}d_{3s}+H.c.)+U_{i}d^{\dag}_{i\uparrow}d_{i\uparrow}d^{\dag}_{i\downarrow}d_{i\downarrow},\label{Eq3}
\end{equation}
where $d^{\dag}_{is}$ ($d_{is}$) creates (annihilates) an electron in the $i$th QD with energy level $\varepsilon_{i}(t)=\varepsilon_{i}^{0}-eW_{D}\cos(\omega t)$; $\varepsilon_{i}^{0}$ is the single particle energy in the $i$th QD. $t$ is the coupling between the QD$_{2}$ and the QD$_{3}$ and $U_{i}$ describes the Coulomb repulsion energy of the $i$th QD.

The last term in Eq. (1), $H_{T}$ , describes electron tunneling between the QDs and leads:
\begin{equation}
  H_{T}=\sum_{k,s,\beta,i=1,2}t_{\beta is}\alpha^{\dag}_{\beta is}d_{is}+H.c.,\label{Eq4}
\end{equation}
where $t_{\beta is}$ represents the QDs-lead coupling.
\

According to Ref. 28, the Rashba SOC has two main effects in a QD system: (1) an extra spin-dependent phase factor appears in the tunneling matrix, and (2) interlevel spin-flip can be induced by Rashba SOC, but not intralevel spin-flip. To simplify the calculation, we assume that each QD has the same energy level. Thus only the first of these two effects is taken into consideration in the present work. In order to simplify the analysis of the self-energies, we use the wide-band limit (WBL), which is an approximation. The energy dependence of the coupling between the leads and the QDs can be neglected by using the WBL. In the WBL, we can use the bandwidth functions to express the retarded self-energy:
\begin{equation}
  \Sigma_{\beta s}^{r}(t,t')=-\frac{i}{2}\delta(t-t')\Gamma_{s}^{\beta},\label{Eq5}
\end{equation}
where $\Gamma_{sij}^{\beta}(\varepsilon,t,t')=2\pi\rho_{\beta}t_{\beta,i}t_{\beta,j}^{*}\textrm{exp}\{i\int_{t'}^{t}W_{\beta}(\tau)d\tau\}$.  Here $\rho_{\beta}$ describes the spin density of states in the $\beta$ lead for spin channel $s$. Therefore, we can use the general time-dependent current method proposed by Wingreen, Jauho, and Meir\cite{10} and obtain the time-dependent current $I(t)$ ($\hbar=1$):
\begin{eqnarray}
I_{\beta s}(t)&=&-2e\textrm{Im}\int_{-\infty}^{t}dt'\int\frac{d\varepsilon}{2\pi}Tr\{e^{-i\varepsilon(t'-t)}\nonumber\\ &&\times\Gamma_{s}^{\beta}(\varepsilon,t,t')[G_{s}^{<}(t,t')+f_{\beta}(\varepsilon)G_{s}^{r}(t,t')]\},\label{Eq6}
\end{eqnarray}
in which $f_{\beta}(\varepsilon)$ is the Fermi distribution function of electrons in the $\beta$ lead. Both the retarded and lesser Green's functions are required. The retarded Green's function $G^{r}$ of the QD is obtained from the corresponding Green's function of the QDs using the Dyson equation:
\begin{equation}
G_{s}^{r}(t,t')=\int\frac{d\varepsilon}{2\pi}\textrm{exp}[-i\varepsilon(t-t')-i\int_{t'}^{t}d\tau W_{D}cos(\omega\tau)]G_{s}^{r}(\varepsilon),\label{Eq7}
\end{equation}
\begin{equation}
G_{s}^{r}(\varepsilon)=[g_{s}^{r-1}(\varepsilon)-\Sigma_{s}^{r}(\varepsilon)],\label{Eq8}
\end{equation}
where $g_{s}^{r}(\varepsilon)$ can be obtained from the Fourier transformation of $g^{r}_{ii}(t,t')=-i\theta(t-t')e^{-i\int_{t'}^{t}\varepsilon_{i}(t_{1})dt_{1}}$. The quantity $n_{is}$ is the average occupation number, and can be calculated using the self-consistent values of $n_{is}$: $n_{is}=\textrm{Im}<G_{iis}^{<}(t,t)>$. As for the lesser Green's function $G^{<}$, we use the Keldysh relation, $G^{<}=G^{r}\Sigma^{<}G^{a}$, which can be easily calculated when $G^{r}$ is known. Using Eq.(7) and Eq. (8), Eq. (6) is reduced to the form
\begin{eqnarray}
I_{\beta s}(t)&=&-e\int\frac{d\varepsilon}{2\pi}\textrm{Im}\{2f_{\beta}(\varepsilon)\Gamma_{s}^{\beta}A_{\beta s}(\varepsilon,t)\nonumber\\
&&+i\Gamma_{s}^{\beta}\sum_{\alpha=L,R}f_{\alpha}(\varepsilon)A_{\alpha s}(\varepsilon,t)\Gamma_{s}^{\alpha}A_{\alpha s}^{\dag}(\varepsilon,t)\},\label{Eq9}
\end{eqnarray}
where
\begin{eqnarray}
A_{\beta s}(\varepsilon,t)&=&\textrm{exp}[i(eW_{\beta}-eW_{D}\sin(\omega t)/\omega)]\nonumber\\
&&\times\sum_{n}J_{n}(e\frac{W_{D}-W_{\beta}}{\omega})e^{in\omega t}G_{s}^{r}(\varepsilon_{n}).\label{Eq10}
\end{eqnarray}
Here, $J_{n}$ is Bessel function and $\varepsilon_{n}=\varepsilon-n\omega$. Eq. (9) is the expression for the instantaneous current. However, experimentally the average current is more relevant. The time average of Eq. (9) is
\begin{widetext}
\begin{equation}
<I>=2e\int\frac{d\varepsilon}{2\pi}\sum_{n}Tr\{[J_{n}^{2}(e\frac{W_{D}-W_{L}}{\omega})f_{L}(\varepsilon)-J_{n}^{2}(e\frac{W_{D}-W_{R}}{\omega})f_{R}(\varepsilon)]\Gamma_{s}^{L}G_{s}^{r}(\varepsilon_{n})\Gamma_{s}^{R}G_{s}^{a}(\varepsilon_{n})\}.\label{Eq11}
\end{equation}
\end{widetext}
\section{Results and discussions}
\subsection{Parallel double dots ($t=0$)}
While the transport properties of single and double QDs have been well studied,\cite{15} for completeness and for later discussions, we analyse parallel double QDs without a side-coupled third QD in this subsection. We used $\omega$ as the units of measurement, and assumed that $\Gamma_{1}^{\beta}=\Gamma_{2}^{\beta}=\Gamma$. Using the Eq. (11) above, the spin-dependent average current of the model can be numerically simulated. In our calculation, we found that the case in which external MW fields are applied symmetrically ($W_{L}=W_{R}$) on the leads is same as the case in which the external MW fields are applied directly to the QDs,\cite{21}. Therefore we take $W_{D}=0$ in our discussion. We begin discussion of the double QDs with both the Rashba SOC and the magnetic field considered in a symmetic ac field. For this situation, Fig. 2 shows the spin-dependent average currents as a function of the energy level of the QDs.
\begin{figure}
\includegraphics[scale=0.6]{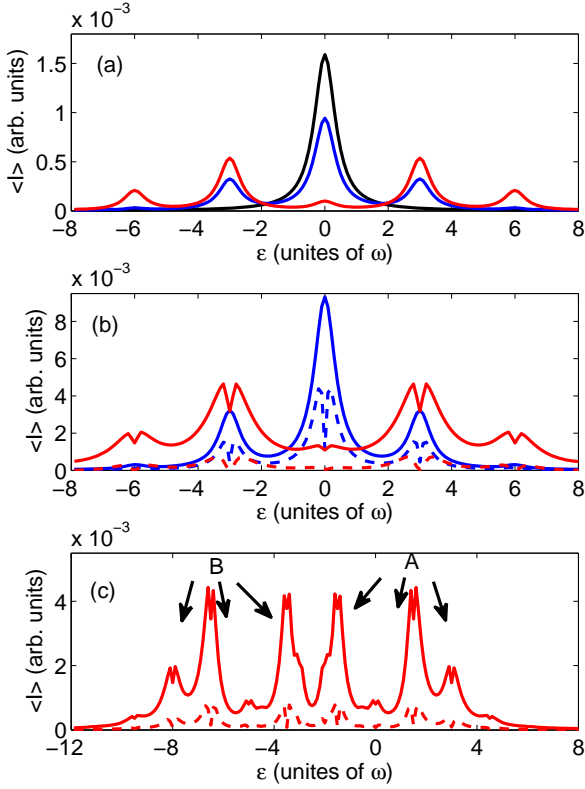}
\caption{\label{fig_2}Spin-dependent average currents, $<I>_{\uparrow}$ (solid line) and $<I>_{\downarrow}$ (dashed line), for QDs in parallel as a function of the electron energy level $\varepsilon$ in the QDs under ac bias with $eW_{L,R}=0$ (black line), $eW_{L,R}=3$ (blue line) and $eW_{L,R}=6$ (red line) (a) without Rashba SOC, magnetic field or $e-e$ interaction when $\hbar\omega=1.5$ (b) with the Rashba SOC $\sigma_{R}=\pi/4$ and magnetic flux $\varphi=\pi/4$ (blue line), $\sigma_{R}=\pi/2$ and $\varphi=3\pi/4$ (red line), for $\hbar\omega=1.5$ and no $e-e$ interaction, (c) with Rashba SOC ($\sigma_{R}=\pi/2$), magnetic flux ($\varphi=3\pi/4$), for $\hbar\omega=1.5$ and $U=5$. The other parameters are: $k_{B}T=0.001$ and $V=0.1$.}
\end{figure}

As illustrated in Fig. 2(a), the dc current (black solid line) has a Lorentzian line shape whose width is determined by $\Gamma$. The peak of the average current occurs at $\varepsilon=0$. There is no spin splitting when the system does not have either Rashba SOC or a magnetic field. In addition, when a harmonic ac source with amplitude $eW_{L,R}/\hbar\omega=1$ (blue solid line) is applied, changes caused by the ac field can be discerned. The current shows two polar values at $\varepsilon=\pm\hbar\omega$. The photo-assisted features are more clearly seen when $eW_{L,R}/\hbar\omega=2$ (red solid line)in which case, there are side peaks located at $\varepsilon=\pm\hbar\omega$ and $\varepsilon=\pm2\hbar\omega$. These peaks are due to the PAT or sideband effect, and each term in the summation of Eq. (11) can be regarded as the contribution of the $n$-photon process.
\begin{equation}
\varepsilon=n\hbar\omega,~~~~n=0,\pm1,\pm2,\ldots.\label{Eq12}
\end{equation}
The average current, shown in Fig. 2(a), is symmetric about $\varepsilon=0$ when the dc source-drain voltage $V=\mu_{L}-\mu_{R}=0.1$. From the PAT point of view, the part of average current with $\varepsilon<0$ is associated with first absorption and then emission of photons, while the part for $\varepsilon>0$ is associated with first emission and then absorption i.e. the time reversed counterpart to $\varepsilon<0$. In addition, the ccentral peak at $\varepsilon=0$ is suppressed which is due to the prefactor $J_{n}^{2}(e\frac{W_{D}-W_{L}}{\omega})f_{L}(\varepsilon)-J_{n}^{2}(e\frac{W_{D}-W_{R}}{\omega})f_{R}(\varepsilon)$ in each term of the summation in Eq. (11), which causes the peak heights for resonant tunneling to become lower for larger $n$. It can be obtained in our calculation that the sum of the heights of all peaks is equal to the height of the original peak.
\

In the Fig. 2(b), we show the spin-dependent average current when the SOC and magnetic field are both included. The results show that the spin-dependent average current $<I>$ of the two spin channels are equal in the absence of both SOC and a magnetic field, but become quite different when they are included. There are two remarkable features in the average current characteristics that arise due to Rashba SOC and the magnetic field. First, the current has a minimum at $\varepsilon=n\hbar\omega$ ($n=0,\pm1,\pm2,\ldots$). We note that the magnitude of the average current can be controlled by $W_{\beta}$. Due to the cophase wave functions along the two paths when the Rashba SOC phase $\sigma_{R}=0$, the current is maximum at $\varepsilon=n\hbar\omega$. As shown in Fig. 2(b) (blue line), when the magnetic field phase is $\varphi=\pi/4$ and the SOC phase is $\sigma_{R}=\pi/4$, all peaks of the spin down current, including the main peaks and sideband peaks, are split at $\varepsilon=n\hbar\omega$. However this splitting phenomenon does not occur in the spin up current [see Fig. 2(b)]. The reason is that the sum of the phases is $-s\sigma_{R}+\varphi=0$ for the spin up current ($s=1$), but the sum is $-s\sigma_{R}+\varphi=\pi/2$ for the spin down current ($s=-1$). The Rashba SOC behaves like a momentum dependent magnetic field which is perpendicular to the system. This effective magnetic field induces a spin-dependent phase difference between the electrons traveling clockwise and counterclockwise between QD$_{1}$ and the QD$_{2}$. Because of the interference between the wave functions along the two paths when $\sigma_{R}\neq0$, the current shows a large decrease at $\varepsilon=n\hbar\omega$ for the spin-down channel. As a result, the spin-down channel peaks are split at $\varepsilon=n\hbar\omega$.
\

The second aspect to note, and the one on which we will concentrate in the following, is that we can obtain 100\% spin polarized current for appropriate values of $\varphi$ and $\sigma_{R}$. For example, when $\varphi=3\pi/4$ and $\sigma_{R}=\pi/2$, we can note two cases with zero current at $\varepsilon=\pm\hbar\omega$ in the spin-down channel. See Fig. 2(b) (red line). In contrast, two non-zero currents occur at the same locations for the spin-up electrons. Thus, a spin-up current can be obtained at $\varepsilon=\pm\hbar\omega$. If $\varphi=-3\pi/4$ and $\sigma_{R}=\pi/2$ or $\varphi=3\pi/4$ and $\sigma_{R}=-\pi/2$ the conductance curves for the spin-up and spin-down channels are interchanged and a spin-down current can be obtained for $\varepsilon=\pm\hbar\omega$. Therefore, the magnitude and direction of the spin polarization of each spin channel can be controlled by the phase factors $\varphi$ and $\sigma_{R}$, which is useful for designing a spin filter under ac bias.
\

We can also note that there are two kinds of peak in Fig. 2(c) which is distinguished by having a non-zero $e-e$ interaction. The A-type peaks are the general PAT peaks discussed above for both spin directions and are a distance $\varepsilon=\pm n\hbar\omega$ away from the main peak at $\varepsilon=0$. The `B' peaks are also a kind of PAT peaks but their positions are modified due to the $e-e$ interaction. The first `B' peak is located at $\varepsilon=-U+\hbar\omega$. The electronic states in the QDs are occupied with energy $\varepsilon=-U$, and peaks at $\varepsilon=-U\pm n\hbar\omega$ produces as the result of a PAT process based on a QD with this energy. In this case, due to the $e-e$ interaction, a photon must have an energy $\varepsilon=-U\pm n\hbar\omega$ in order for it to be absorbed or emitted.
\begin{figure}
\includegraphics[scale=0.6]{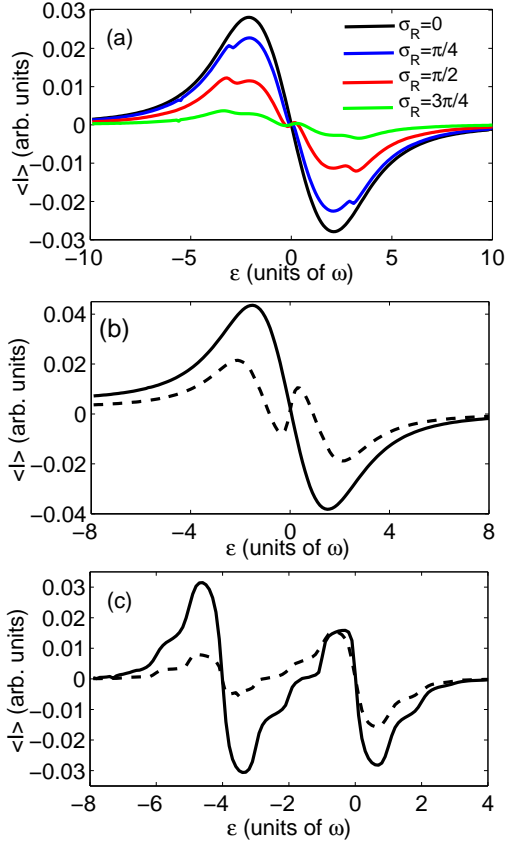}
\caption{\label{fig_3}For the asymmetric case ($eW_{L}=0,eW_{R}=2$), the spin-dependent average currents, $<I>_{\uparrow}$ (solid line) and $<I>_{\downarrow}$ (dashed line), are shown as a function of the electron energy level $\varepsilon$ of the QDs under an asymmetric ac bias (a) with different strengths of Rashba SOC, $\varphi=0$, $U=0$ and $\hbar\omega=3$, (b) with $\varphi=\pi/4$, $\sigma_{R}=\pi/4$, $U=0$ and $\hbar\omega=0.8$ (c) with $\varphi=\pi/4$, $\sigma_{R}=\pi/4$, $U=4$ and $\hbar\omega=2$. The other parameters are: $k_{B}T=0.001$ and $V=0$.}
\end{figure}

Fig. 3(a) shows that the average current $<I>$ versus the intradot energy $\varepsilon$ when the system is subject to an asymmetrical time-dependent external field ($W_{L}=0$ and $W_{R}\neq0$) for $\sigma_{R}=0$, $\pi/4$, $\pi/2$ and $3\pi/4$. When $\sigma_{R}=0$, the main peak is located at $\varepsilon=\Gamma^{L(R)}$. For $\sigma_{R}\neq0$, the subsidiary peak exceeds the main peak (near $\varepsilon=\pm\hbar\omega$) and can not be neglected. The shoulder on the left side of the main resonant peak and a negative current on the right side in Fig. 3(a), result from the electron-photon pump. With increasing $\sigma_{R}$, the magnitude of the average current is reduced and the shoulder becomes clearer, but the location of the PAT peaks is independent of the strength of the Rashba SOC. The distance between the PAT peaks and the point $\varepsilon=0$ is almost unchanged, and is equal to $\varepsilon=\pm\hbar\omega$. \

It should be noted that when both $\varphi$ and $\sigma_{R}$ are taken into account, the current is spin polarized and can be controlled. In Fig. 3(b), when the system is subject to an asymmetric time-dependent field, if $-1<\varepsilon<0$, the spin up current $<I>_{\uparrow}$ is positive, which means that the spin up current $<I>_{\uparrow}$ flows along the positive direction (left to right), while the spin down current $<I>_{\downarrow}$ flows in the negative direction (right to left). In the range $0<\varepsilon<1$, however, the opposite situation occurs. This means that the current is spin polarized and the polarization can be controlled by the magnetic field and Rashba SOC in the asymmetric MW situation. Here we should point out that since we considered the intradot Coulomb interaction in our model, we obtain a series of Coulomb oscillation shoulders in Fig. 3(c) located at $\varepsilon=-U$ and $\varepsilon=-U\pm\hbar\omega$.
\begin{figure}
\includegraphics[scale=0.48]{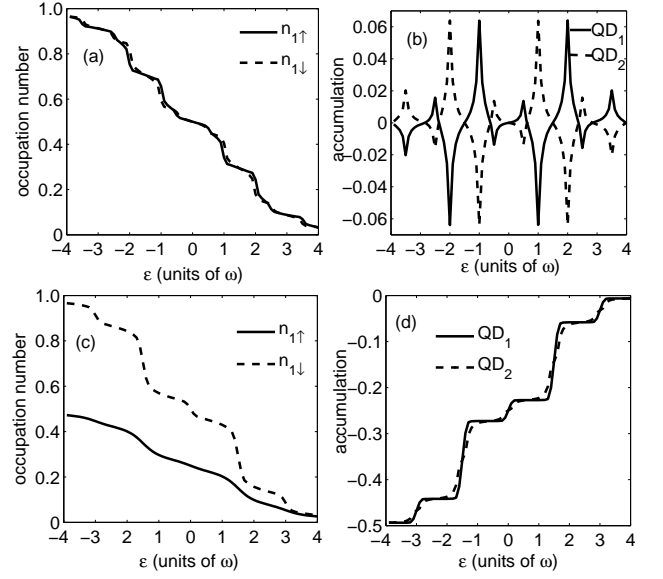}
\caption{\label{fig_4} (a) Spin occupation numbers $n_{1\uparrow}$, $n_{1\downarrow}$ and (b) accumulations $\Delta n_{1}$, $\Delta n_{2}$ versus electron energy $\varepsilon$ in QDs when $\sigma_{R}=\pi/4$, $\varphi=0$, $V=1$ and $eW_{L,R}=2\hbar\omega=3$. (c) Spin occupation numbers $n_{1\uparrow}$, $n_{1\downarrow}$ and (d) accumulations $\Delta n_{1}$, $\Delta n_{2}$ versus electron energy $\varepsilon$ in QD$_{1}$ when $\sigma_{R}=\pi/4$, $\varphi=\pi/4$, $V=0.2$ and $eW_{L,R}=2\hbar\omega=3$. The other parameters are: $k_{B}T=0.001$ and $U=0$.}
\end{figure}

Spin polarization of the average current can not realized by Rashba SOC alone. [See Fig. 3(a)] However the occupation number can be polarized by the SOC alone. To make clear the effect of Rashba SOC in our model, we introduce the total effective coupling strength $T_{L1}$ between the QDs (e.g. QD$_{1}$) and the left lead:
\begin{equation}
T_{L1s}=|t_{L1s}+t_{L2s}g_{22}^{r}t_{R2s}(-i\pi\rho)t_{R1s}e^{-is\sigma_{R}}|^{2},\label{Eq13}
\end{equation}
\begin{equation}
T_{R1s}=|t_{R1s}e^{-is\sigma_{R}}+t_{R2s}g_{22}^{r}t_{L2s}(-i\pi\rho)t_{L1s}|^{2}.\label{Eq14}
\end{equation}

Due to the fact that $\sigma_{R}\neq0$, we can see $T_{Lis}\neq T_{Ris}$ which causes spin accumulation($\Delta n_{i}=\Delta n_{i\uparrow}-\Delta n_{i\downarrow}$) in the QDs. The spin up and spin down occupation numbers and the spin accumulation versus the intradot energy level $\varepsilon$ in the QDs are shown in Fig. 4(a) and Fig. 4(b) respectively for $\sigma_{R}=\pi/4$ when the system is under a symmetric time-dependent external field ($eW_{L,R}=2\hbar\omega=3$). Though the current $<I>$ is not spin polarized, the spin occupation number $n_{i\uparrow}$ is not equal to $n_{i\downarrow}$ when the Rashba SOC is considered. As a result, the intradot spin accumulation $\Delta n_{i}$ is non-zero. The spin accumulation in QD$_{1}$ is opposite to that in QD$_{2}$ [see Fig. 4(b)]. At $\varepsilon=n\hbar\omega$, the spin accumulation $\Delta n_{1}=\Delta n_{2}=0$. In the vicinity of the $\varepsilon=n\hbar\omega$, however, the spin accumulation $\Delta n_{i}$ has a maximum value, which leads to a large polarization of one QD. However the spin accumulation of the system as a whole is zero for any $\varepsilon$, with the result that a net spin polarization does not form in double QD systems. Even for a small $\sigma_{R}$ and dc bias $V$, the two QDs have polarizations with opposite signs. This enables us to control the spin accumulation using Rashba SOC. It would appear that the production of spin occupation and accumulation should be experimentally feasible with present nanotechnology.
\

The spin precession angle can be described as $\sigma_{R}=\alpha_{R}m^{\ast}L/\hbar^{2}$, and the strength of the Rashba SOC is about $3\times10^{-11}$eVm, which can be controlled experimentally. Here $L$ is the size of the QD. The magnitude of $\sigma_{R}$ can reach yet larger values experimentally when the dimension of the QD is about $100$ nm and $m^{\ast} = 0.036m_{e}$.\cite{29} Fig. 4(c) and Fig. 4(d) describe the dependence of the intradot spin occupation numbers and the spin accumulation on the energy level $\varepsilon$ in the QDs for the case $\sigma_{R}=\pi/4$ and $\varphi=\pi/4$. Spin accumulation in the QDs presents a `step' shape when $\varphi$ and $\sigma_{R}$ are both considered [see the Fig. 4(d)]. This is quite different from the situation in Fig. 4(b) where only $\sigma_{R}$ is considered. It may be seen that the width of the `steps' is just $\hbar\omega$. Even for a small $\varphi$, $\Delta n$ is large. For example, at $\varepsilon=-4\hbar\omega$, $\Delta n_{i}\approx -0.5$; which is quite large for a spin polarization that relies on small values of $\varphi$ and $\sigma_{R}$ without $e-e$ interactions.
\begin{figure}
\includegraphics[scale=0.55]{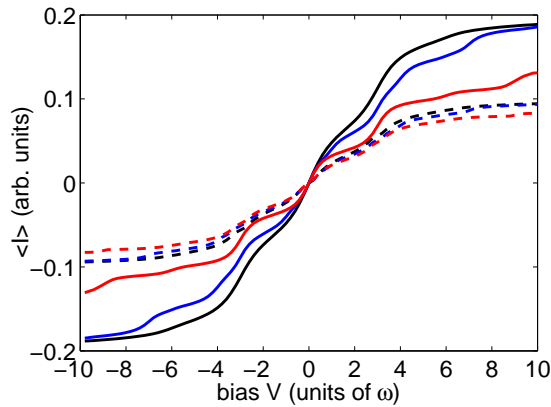}
\caption{\label{fig_5} Spin-dependent average current, $<I>_{\uparrow}$ (solid line) and $<I>_{\downarrow}$ (dashed line), versus dc bias $V$ with a symmetric ac bias ($eW_{L,R}=\hbar\omega=3$) and $U=0$ (black curve), $U=2$ (blue curve) and $U=6$ (red curve). The other parameters are: $k_{B}T=0.001$, $\varepsilon=0$, $\sigma_{R}=\pi/4$ and $\varphi=\pi/4$.}
\end{figure}

We now investigate the effect of the interaction between the electrons in the QD system. The spin-dependent average currents versus the dc bias $V$ for different strengths of the Coulomb interaction are shown in Fig. 5. From Fig. 5, we can see that spin polarization of the average current indeed occurs in the QDs with a finite dc bias $V$ in the ac field. When the bias $V=0$, the average current for both the spin-up and spin-down channels should be zero. The magnitude of the spin polarization increases when the dc bias $V$ increases, and the spin polarized direction can be reversed by reversing the dc bias. Thus the direction and the magnitude of the spin polarization are easily regulated by the dc bias in a symmetric ac field. Fig. 5 also illustrates that the change in the spin-down current in our model is tiny for any value of $U$ and any dc bias voltage, whereas the spin-up current has a finite value and decreases as $U$ increases. Repulsion between electrons with spin-up or spin-down results from the Coulomb interaction, $U$, which leads to a reduction in the magnitude of spin polarization. The shoulders that appear at $V=\pm3$ and $V=\pm6$ are attributed to the PAT effect in the symmetric ac field.

\begin{figure}
\includegraphics[scale=0.6]{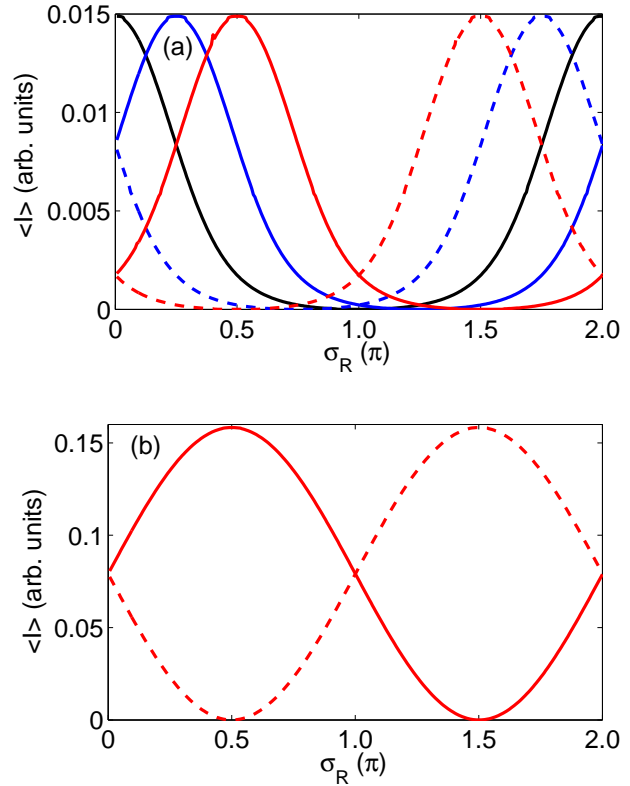}
\caption{\label{fig_6} Spin-dependent average current, $<I>_{\uparrow}$ (solid line) and $<I>_{\downarrow}$ (dashed line), versus the strength of Rashba SOC $\sigma_{R}$ under a symmetric ac bias ($eW_{L,R}=2\hbar\omega=2$) with (a) $\varphi=0$ (black curve), $\varphi=\pi/4$ (blue curve), $\varphi=\pi/2$ (red curve) and $V=0.1$ (b) $\varphi=\pi/2$ (red curve) and $V=3$ . The other parameters are: $k_{B}T=0.001$, $\varepsilon=0$ and $U=0$.}
\end{figure}

Next we study how the spin-dependent currents change with the strength of the Rashba SOC, $\sigma_{R}$. The average currents $<I>$ versus the $\sigma_{R}$ are illustrated in figure 6 from which we can see that the value of $<I>$ is sensitive to the spin-dependent phase $\sigma_{R}$. The period of the time-averaged current is 2$\pi$. From Fig. 6(a), we can also note that the average current is not polarized when only Rashba SOC is considered (black curve). This result is the same as shown in Fig. 3(a). As the magnetic flux $\varphi$ increases, the spin-up current and spin-down current gradually separate. When $\varphi=\pi/2$, the polarization is as large as 100\% for the given set of system parameters ($\sigma_{R}=\pi/2$ or $\sigma_{R}=3\pi/2$). In the situation where $\varphi=\pi/2$, the transmitted electrons in the spin-up channel can undergo constructive interference in the double QD system. However, at the same time, the spin-down electrons undergo destructive interference, which results in the maximum of spin polarization. Therefore, a purely spin-up current or spin-down current can be chosen by adjusting the strength of the Rashba SOC for $\varphi=\pi/2$ case. In addition, there is no spin polarization when $\sigma_{R}=2n\pi$. Thus, the spin polarized current can be regulated by the phase induced by Rashba SOC in an ac field. With increasing dc bias $V$, the magnitude of the spin polarization also increases significantly, as can be see in Fig. 6(b).

\subsection{ Parallel double dots with a side-coupled dot ($t\neq0$)}
\begin{figure}
\includegraphics[scale=0.45]{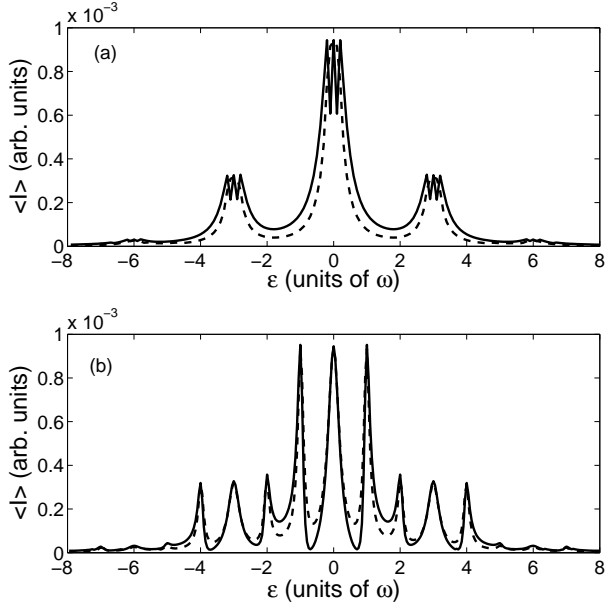}
\caption{\label{fig_7} Spin-dependent average current, $<I>_{\uparrow}$ (solid line) and $<I>_{\downarrow}$ (dashed line), versus electron energy $\varepsilon$ under symmetric ac bias ($eW_{L,R}=\hbar\omega=3$) with (a) $t=\Gamma=0.2$ and (b) $t=5\Gamma=1$. The other parameters are: $k_{B}T=0.001$, $V=0.1$, $\sigma_{R}=\pi/4$, $\varphi=\pi/4$ and $U=0$.}
\end{figure}

The effect of the coupling term $t$ between the QD$_{2}$ and the QD$_{3}$ on the spin-dependent average current through the system described above is illustrated in Fig. 7. The spin-dependent average current in the case $t=\Gamma$ is quite different from the $t=0$ case shown in Fig. 2(b) (blue line). All the peaks of the spin up current, including the main peaks and sideband peaks, are split at $\varepsilon=n\hbar\omega$ and the split peaks of the spin down current disappear. From Eq. (11) we find that the split peaks are located at
\begin{equation}
\varepsilon=n\hbar\omega\pm\sqrt{t^{2}-\Gamma^{2}\sin^{2}(\frac{\varphi-s\sigma_{R}}{2})},n=0,\pm1,\pm2,\ldots.\label{Eq15}
\end{equation}
These peaks all result from the PAT effect due to the coupling between QD$_{2}$ and QD$_{3}$, and each term in the sum in Eq. (11) can be viewed as the contribution from the coupling for $n$-photon processes. It should be noted that if $t=\Gamma$ and $\varphi=\sigma_{R}=\pi/4$, the peaks for the spin-down channel should appear at $\varepsilon=n\hbar\omega$, which is the position for conventional PAT peaks. It can be seen in Fig. 7(a) that the shape of the spin-up peaks, $\varepsilon=n\hbar\omega\pm \sqrt{t^{2}-\frac{1}{2}\Gamma^{2}}$, is distinctly different from that of the spin-down peaks, which correspond to another kind of sideband peak. Due to quantum interference between the two paths for transmitted electrons (input lead-QD$_{1}$-output lead) and (input lead-QD$_{2}$-output lead), the conductance spectrum shows prominent peaks in Fig. 2. The effect of the coupling between QD$_{2}$ and QD$_{3}$ is to modify the above quantum interference between the QD$_{1}$ path and QD$_{2}$ path. When $t=\Gamma$, the influence of coupling term has only a very small role. The appearance of the resonance peaks at $\varepsilon=n\hbar\omega\pm \sqrt{t^{2}-\frac{1}{2}\Gamma^{2}}$ is due to the path of transmitted electron being from the input lead, through QD$_{2}$-QD$_{3}$-QD$_{2}$ to the output lead. A new quantum state can be formed by QD$_{2}$ and QD$_{3}$. When the coupling strength, $t$, between QD$_{2}$ and QD$_{3}$ increases, the change in the average current can be seen with significant modification of the quantum interference through the triple QD system. Therefore a resonance band forms when the strength of coupling $t$ is similar to $\Gamma$. The three PAT peaks thus occur in the average current as the value of $t$ increases. Three quantum states can be formed by the triple QD, and the three PAT peaks can be seen clearly in Fig. 7(b). In Fig. 7(b), with $t=5\Gamma$ and $\varphi=\sigma_{R}=\pi/4$, the spin up peaks and spin down peaks appear at approximately $\varepsilon=n\hbar\omega\pm t$ and $\varepsilon=n\hbar\omega$. It may be seen that the three corresponding PAT peaks for both spin channels can be clearly distinguished only when $t\gg\Gamma$. The PAT peaks due to the coupling between QD$_{2}$ and QD$_{3}$ decrease when $t$ is similar to $\Gamma$. That is, the PAT peaks induced by the coupling between QD$_{2}$ and QD$_{3}$ becomes smaller as $t$ decreases. The three PAT peaks can not be distinguished in the $t<\Gamma$ case.
\begin{figure}
\includegraphics[scale=0.45]{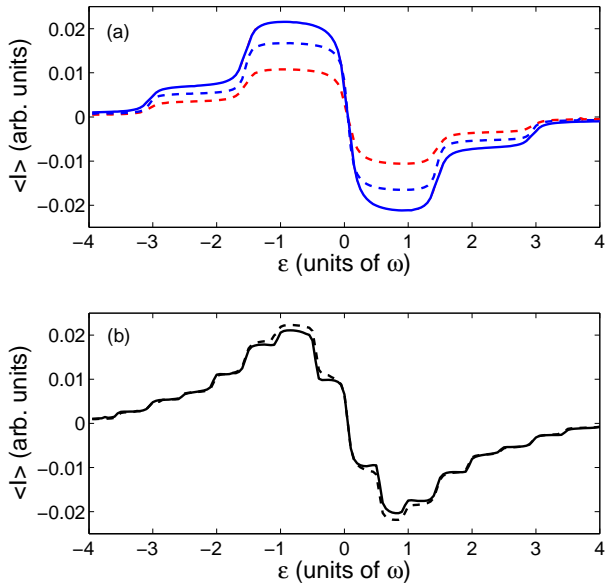}
\caption{\label{fig_8} Spin-dependent average currents, $<I>_{\uparrow}$ (solid line) and $<I>_{\downarrow}$ (dashed line), versus electron energy $\varepsilon$ for two QDs connected in parallel with a side-coupled QD, under an asymmetric ac bias ($eW_{L}=0$, $eW_{R}=2\hbar\omega=3$)with (a) $t=0$ (red line), $t=\Gamma=0.05$ (blue line) and (b) $t=10\Gamma=0.5$ (black line). The other parameters are: $k_{B}T=0.001$, $V=0$, $\sigma_{R}=\pi/4$, $\varphi=\pi/4$ and $U=0$.}
\end{figure}

Fig. 8(a) shows a numerical calculation of the spin-dependent average current $<I>$ versus $\varepsilon$ under asymmetric ac bias. For the $t=0$ case, a shoulder occurs on the left side of the main peak, and a negative current shoulder appears on the right side of the main peak. The negative current shoulder and the positive current shoulder in the curve are due to the electron-photon pump effect. With increased coupling such as $t=\Gamma$ (weak-coupling), changes in the spin-up channel caused by the coupling are too small to discern. However, the shoulder for spin down electrons is higher than for the $t=0$ case. The height of the shoulder is determined by the coupling. However, the location of the PAT shoulder is independent of the MW field amplitude and the coupling strength.

\
To examine the effects of strong coupling, we set $t=10\Gamma$, and use $W_{L}=0$, $W_{R}\neq0$, with the source-drain voltage $V=\mu_{L}-\mu_{R}=0$. In this case the transmitted electrons flowing in the left lead will be free of the MW fields, and only the transmitted electrons in the right lead feel the MW fields. The electron-photon pump effect appears as steps in the $<I>$ versus $\varepsilon$ curves as shown in Fig. 8(b). The locations of the shoulders due to the electron-photon pump are not only at $n\hbar\omega$ but also at $n\hbar\omega\pm t$. This means that the tunneling electron can emit or absorb photons with different frequencies in a more complicated way. For example, the electron for the $\varepsilon=0$ energy level can absorb or emit a photon of energy $n\hbar\omega$ in the usual manner. However, an electron in the $\varepsilon=\pm t$ energy level may also absorb or emit a photon of energy $n\hbar\omega$ with the result that coherence effects arise. The result is a more complicated multiple-PAT effect arising from the interplay of the new quantum state formed by the coupling between QD$_{2}$ and QD$_{3}$, and the electron-photon pump. In fact the MW field and the coupling considered in the system cause the new quantum state corresponding to the three QDs to participate in the transmission, and the MW field applied on the right lead in an asymmetric way induces the electron-photon pump effect.

\section{\label{sec:level4}Conclusions}
In summary, we have studied the PAT effect and the electron-photon pump effect through two QDs connected in parallel with a side-coupled QD, the whole system being irradiated by a MW field. The spin-dependent average currents $<I>_{\uparrow}$, $<I>_{\downarrow}$, the spin occupation $n_{is}$ and spin accumulation $\Delta n_{i}$ were obtained utilizing the Keldysh nonequilibrium Green's function method. When only the Rashba SOC is considered, spin polarization can be produced in the QDs, and can be seen in the spin accumulation. When we consider the combined effect of both Rashba SOC and a magnetic flux, both the intradot occupation numbers and the time averaged current through the system are polarized. A pure spin polarized current can be generated due to the nonzero spin-dependent phase $\sigma_{R}$ and the magnetic flux $\varphi$ in the presence of an ac bias. This provides an efficient way to generate a pure spin polarization current in nanostructures. When QD$_{2}$ and QD$_{3}$ are coupled, several interesting effects related to the more complicated level structure of the QDs are expected to occur. In particular, the multiple-photon-assisted tunneling is more complicated and a new kind of PAT peak obtained by controlling the strength of the coupling arises. The model considered here can be realized using present technologies. These results are expected to be useful for device design and quantum computation in the future.
\begin{acknowledgments}
This work was supported by the National Natural Science Foundation of China (Grant Nos. 61176089 and 11104059), the Natural Science Foundation of Hebei Province (Grant Nos. A2011205092 and A2011208010). We are also very grateful to Professor N. E. Davison for the enhancement of the English writing.
\end{acknowledgments}


\nocite{*}
\begin{thebibliography}{}
\bibitem{1}P. K. Tien, and J. P. Gorden, Phys. Rev. \textbf{129}, 647 (1963).
\bibitem{2}A. D. Stone, M. Y. Azbel, and P. A. Lee, Phys. Rev. B \textbf{31}, 1707 (1985).
\bibitem{3}D. Sokolovski, Phys. Rev. B \textbf{37}, 4201 (1988).
\bibitem{4}H. C. Liu, Phys. Rev. B \textbf{43}, 12538 (1991).
\bibitem{5}P. Johansson, Phys. Rev. B \textbf{41}, 9892 (1990).
\bibitem{6}P. Johansson, and G. Wendin, Phys. Rev. B \textbf{46}, 1451 (1992).
\bibitem{7}Y. V. Nazarov, Physica B \textbf{189}, 57 (1993).
\bibitem{8}C. Bruder, and H. Schoeller, Phys. Rev. Lett. \textbf{72}, 1076 (1994).
\bibitem{9}X. Chen, D. Liu, W. Duan, and H. Guo, Phys. Rev. B \textbf{87}, 085427 (2013).
\bibitem{10}N. S. Wingreen, A. P. Jauho, and Y. Meir, Phys. Rev. B \textbf{48}, 8487 (1993).
\bibitem{11}A. P. Jauho, N. S. Wingreen and Y. Meir, Phys. Rev. B \textbf{50}, 5528 (1994).
\bibitem{12}L. Y. Chen, and C. S. Ting, Phys. Rev. B \textbf{43}, 2097 (1991).
\bibitem{13}E. Runge, and H. Ehrenreich, Phys. Rev. B \textbf{45}, 9145 (1992).
\bibitem{14}Q. F. Sun, and T. H. Lin, J. Phys.: Condens. Matter \textbf{9}, 4875 (1997).
\bibitem{15}K. Shibata, A. Umeno, K. M. Cha, and K. Hirakawa, Phys. Rev. Lett. \textbf{109}, 077401 (2012).
\bibitem{16}R. N. Shang, H. O. Li, G. Cao, M. Xiao, T. Tu, H. W. Jiang, G. C. Guo, and G. P. Guo, Appl. Phys. Lett. \textbf{103}, 162109 (2013).
\bibitem{17}K. Wang, C. Payette, Y. Dovzhenko, P.W. Deelman, and J. R. Petta, Phys. Rev. Lett. \textbf{111}, 046801 (2013).
\bibitem{18}T. Obata, M. P. Ladri\`{e}re, Y. Tokura, and S. Tarucha, New J. Phys. \textbf{14}, 123013 (2012).
\bibitem{19}L. P. Kouwenhoven, S. Jauhar, K. McCormick, D. Dixon, P. L. McEuen, Yu. V. Nazarov, N. C. van der Vaart, and C. T. Foxon, Phys. Rev. B \textbf{50}, 2019 (1994).
\bibitem{20}L. P. Kouwenhoven, S. Jauhar, J. Orenstein, P. L. McEuen, Y. Nagamune, J. Motohisa, and H. Sakaki, Phys. Rev. Lett. \textbf{73}, 3443 (1994).
\bibitem{21}Q. F. Sun, and T. H. Lin, Phys. Rev. B \textbf{56}, 3591 (1997).
\bibitem{22}Q. F. Sun, J. Wang, and T. H. Lin, Phys. Rev. B \textbf{61}, 12643 (2000).
\bibitem{23}X. T. An, and J. J. Liu, Phys. Lett. A. \textbf{372}, 6790 (2008).
\bibitem{24}X. T. An, H. Y. Mu, Y. X. Li, and J. J. Liu, Phys. Lett. A. \textbf{375}, 4078 (2011).
\bibitem{25}X. T. An, and J. J. Liu, App. Phys. Lett. \textbf{95}, 163501 (2009).
\bibitem{26}Z. L. He, and T. Q. L\"{u}, Phys. Lett. A \textbf{376}, 2501 (2012).
\bibitem{27}C. Guan, Y. H. Xing, C. Zhang, and Z. S. Ma, Appl. Phys. Lett. \textbf{102}, 163116 (2013).
\bibitem{28}Q. F. Sun, J. Wang, and H. Guo, Phys. Rev. B \textbf{71}, 165310 (2005).
\bibitem{29}F. Mireles, and G. Kirczenow, Phys. Rev. B \textbf{64}, 024426 (2001).
\end{thebibliography}
\end{document}